\documentclass[superscriptaddress,twocolumn, aps,prl, reprint, showpacs]{revtex4-1}

\usepackage{epsfig,graphicx,times}

\usepackage{amssymb}
\usepackage{amsmath}
\usepackage[utf8]{inputenc}
\usepackage{bm}

\begin{document}
\title{Geometric-phase interference in a Mn$_{12}$ single-molecule magnet with four-fold rotational symmetry}

\author{S. T. Adams}
\affiliation{Department of Physics, Amherst College, Amherst, MA 01002-5000}
\author{E. H. da Silva Neto}
\altaffiliation[Current address: ]{Department of Physics, Princeton University, Jadwin Hall, Princeton, NJ 08544}
\author{S. Datta}
\author{J. F. Ware}
\altaffiliation[Current address: ]{Department of Physics,
University of Michigan,
450 Church Street,
Ann Arbor, MI 48109-1040}
\affiliation{Department of Physics, Amherst College, Amherst, MA 01002-5000}
\author{C. Lampropoulos}
\altaffiliation[Current address: ]{University of North Florida, Department of Chemistry,
Building 50, Room 3500,
1 UNF Drive,
Jacksonville, FL 32224-7699}
\author{G. Christou}
\affiliation{Department of Chemistry, University of Florida, Gainesville, FL}
\author{Y. Myaesoedov}
\author{E. Zeldov}
\affiliation{Department of Condensed Matter Physics, The Weizmann Institute of Science, Rehovot, Israel}
\author{Jonathan R. Friedman}
\email[Corresponding author: ]{jrfriedman@amherst.edu}
\affiliation{Department of Physics, Amherst College, Amherst, MA 01002-5000}
\affiliation{Institute for Quantum Computing, University of Waterloo,
Waterloo, Ontario, Canada,
N2L 3G1}

\date{\today}

\begin{abstract}
We study the magnetic relaxation rate $\Gamma$ of the single-molecule magnet Mn$_{12}$-tBuAc as a function of magnetic field component $H_T$ transverse to the molecule's easy axis.  When the spin is near a  magnetic quantum tunneling resonance,  we find that $\Gamma$ increases abruptly at certain values of $H_T$.  These increases are observed just beyond values of $H_T$ at which a geometric-phase interference effect suppresses tunneling between two excited energy levels.  The effect is  washed out by rotating $H_T$ away from the spin's hard axis, thereby suppressing the interference effect.  Detailed numerical calculations of $\Gamma$ using the known spin Hamiltonian accurately reproduce the observed behavior.  These results are the first experimental evidence for geometric-phase interference in a single-molecule magnet with true four-fold symmetry.
\end{abstract}


\maketitle

Geometric-phase effects are responsible for many fascinating phenomena in classical and quantum physics from how a cat rights itself while in free fall to the dynamics of charged particles in electromagnetic fields, e.g. the Aharonov-Bohm effect~\cite{755}.  One formulation of geometric-phase effects involves a path-integral approach in which the interference of paths is modulated by the geometric phase difference between the paths~\cite{149, 150, 166, 757}. Such interference effects can reveal the underlying symmetries of the system's Hamiltonian.  The dynamics of spins provide a natural way to explore quantum geometric phases for, as Berry showed in his pioneering work~\cite{756}, a system near a degeneracy point can be mapped onto a spin in an effective magnetic field.

Single-molecule magnets (SMMs) are an ideal test bed for exploring spin geometric-phase interference.  In these systems, each molecule behaves as a large spin with a well-defined Hamiltonian determined by the symmetry of the molecule and its environment~\cite{748}.
The interactions between molecules in a crystal are typically weak and the sample behaves as an ensemble of nominally identical large-spin objects.  In many SMMs, 
the spins have a large anisotropy barrier separating the preferred ``up'' and ``down'' directions.  This leads to hysteresis and slow relaxation between these easy-axis directions.  A geometric-phase effect can lead to interference between tunneling paths, thus modulating the rate at which spins flip direction.  In a ground-breaking experiment, Wernsdorfer and Sessoli~\cite{162} found oscillations in the probability of magnetization tunneling as a field applied along the hard axis modulated the interference between two tunneling paths.  This observation confirmed a theoretical prediction by Garg~\cite{166} and ignited intense theoretical study of related phenomena~\cite{598, *647, *646, 252, 246, 258, 608, *592, 593, 607, 594, 762}.  Geometric-phase interference between tunneling paths has been observed in other SMMs that have effective two-fold symmetry, where tunneling involves the interference between two equal-amplitude paths~\cite{442, *595}.  It has also been seen in antiferromagnetic molecular wheels~\cite{759} and in SMM dimers~\cite{697, 681}.  Such interference effects in the bellwether SMM Mn$_{12}$Ac are complicated by the presence of solvent disorder~\cite{289}, which breaks the SMM's nominal four-fold symmetry, resulting a competition between second-order and fourth-order anisotropies ~\cite{226,122}. Here we report the observation of a geometric-phase interference effect in [Mn$_{12}$O$_{12}$(O$_2$CCH$_2$-$^t$Bu)$_{16}$(CH$_3$OH)$_4$]$ \cdot $CH$_3$OH
(hereafter Mn$_{12}$-tBuAc), a variant of Mn$_{12}$Ac that is free of solvent disorder and maintains its four-fold rotational symmetry~\cite{571, *569, 760,645,798, christos}.  Unlike previous observations of geometric-phase interference, which involved ground-state tunneling, the interference effect described herein is observed in the thermally assisted tunneling regime where tunneling takes place near the top of the barrier.  The interference effect provides a fingerprint that affords an unprecedented ability to clearly identify which levels participate in the thermally assisted process.

\begin{figure}[tb]
\centering
\includegraphics[width=1\linewidth]{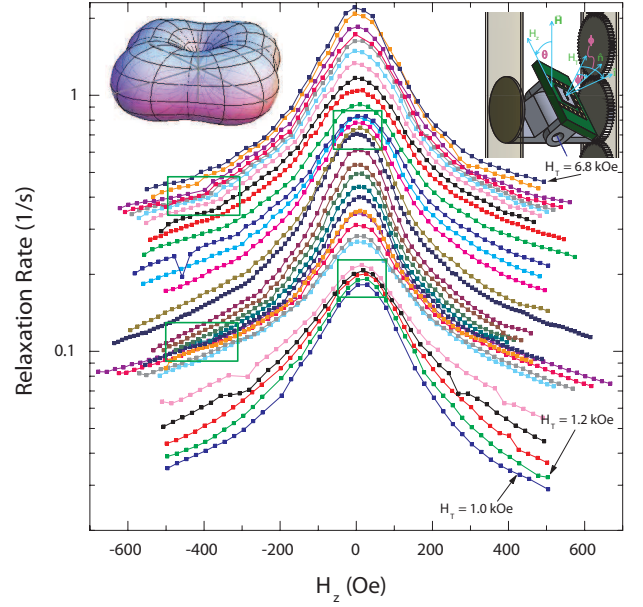}
\caption{(Color online) Measured magnetic relaxation rate as a function of longitudinal field near the zero-field resonance for several values of transverse field (H$_T$) from 1.0 kOe to 6.8 kOe in 200 Oe increments.  The data was taken from sample A at 3.10 K.  The green boxes indicate regions where curves for different values of H$_T$ tend to bunch together. The left inset shows the classical energy landscape in a spherical polar plot for a spin described by Eq.\ref{Mn12Ham}.  The z axis is the easy axis (energy minima) while the x and y axes are the hard axes (maxima).  The value of $C$ has been greatly exaggerated to make the four-fold symmetry evident. The right inset shows a schematic of the apparatus, illustrating control of angles $\theta$ and $\phi$.  $\mathbf{\hat{n}}$ is normal to detector plane; \textbf{\^{i}} is a hard axis direction.}
\label{ratevHz}
\end{figure}

\begin{figure}[tb]
\centering
\includegraphics[width=1\linewidth]{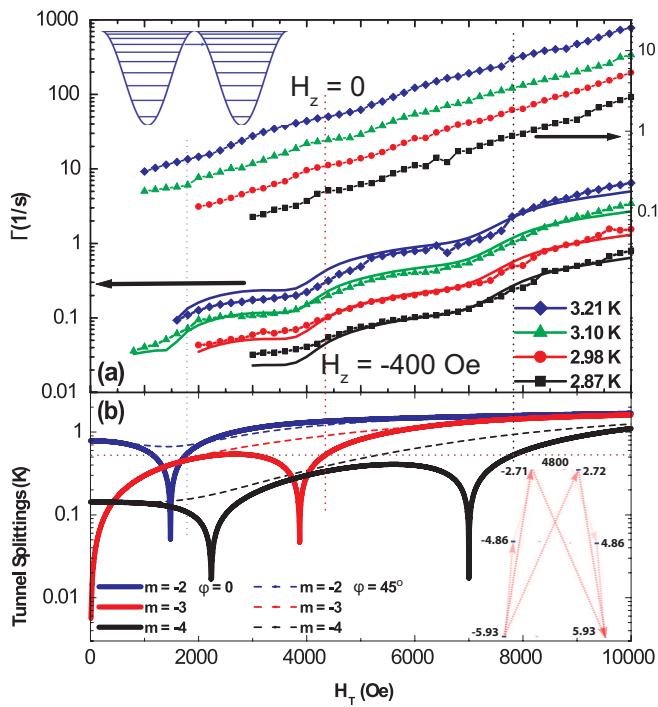}
\caption{(Color online) (a) Relaxation rate as a function of transverse field at $H_z=0$ and $H_z=-400$ Oe, taken from the data shown in Fig.~\ref{ratevHz}.
The solid curves are the results of simulations of the relaxation rate using Eq.~\ref{Mn12tHam} with $\phi=7^\circ$, as described in the text. (b) Calculated tunnel splitting for energy-level pairs $m=\pm 2$, $m=\pm 3$, and $m=\pm 4$, as indicated, for  $\phi=0$ (solid) and $\phi=45^\circ$ (dashed). The dashed horizontal line running through these calculations marks a threshold splitting at which the system makes a transition from one dominant pair of  tunneling levels to another. The dashed vertical lines show the transverse field at which this transition occurs and its correspondence to rapid increases in the relaxation rate. Inset: Probability current diagram for high-lying energy levels at $H_T = 4.8$ kOe.  The numerical labels indicate the expectation values of $S_z$ for the corresponding energy level; the opacity of the arrows indicates the magnitude of the associated current.}
\label{ratevHT}
\end{figure}

In 2002, Park and Garg~\cite{216} and, independently, Kim~\cite{414}, predicted that an interference effect should be observed in SMMs with fourth-order transverse anisotropy, described by the Hamiltonian
\begin{equation}
{\cal H} = - DS_z^2 + (C/2)(S_+^4 + S_-^4) - g\mu_B S \cdot H, \label{Mn12Ham}
\end{equation}
where $S_{\pm}=S_x\pm i S_y$. In zero field, such a spin system has the classical energy landscape shown in the left inset of Fig.~\ref{ratevHz}. Applying a magnetic field $H_T$ along one of the four hard directions ($\pm$ x and $\pm$ y for $C>0$) preserves reflection symmetry through the z-hard plane, allowing for interference between equal-amplitude tunneling paths that virtually pass through the saddle points in the landscape.  This interference induces oscillations in the tunneling probability as a function of $H_T$.  Mn$_{12}$-tBuAc is reasonably well described by the above Hamiltonian with the addition of sixth-order terms consistent with four-fold symmetry~\cite{645}:

\begin{eqnarray}
{\cal H} & = &- DS_z^2 - AS_z^4 - A^\prime S_z^6+ \left(C/2\right)\left(S_+^4 + S_-^4\right) \nonumber \\ & &+\left(C^\prime/2\right)\left(S_z^2\left(S_+^4 + S_-^4\right)+ h.c.\right)\label{Mn12tHam}  \\
& &- \mu_B \left(g_z S_z H_z+ g_\bot\left(S_x cos \phi + S_y sin \phi\right)H_T\right), \nonumber
\end{eqnarray}

\noindent where $D = 0.568$ K, $A=0.69$ mK, $A^\prime=3.3$ $\mu$K, $C= 50$ $\mu$K, $C^\prime=-0.79$ $\mu$K, $g_z = 2.00$ and $g_\bot=1.93$.  $H_z$
is the longitudinal
component of the magnetic field, and $\phi$ measures the angle between $H_T$ and the x axis.  Much of the system's dynamics can be understood in terms of the double-well potential shown in the inset of Fig.~\ref{ratevHT}a, which shows the system's energy as a function of the angle between the spin vector and the easy axis (z) direction.  The spin has $2S+1$ energy levels, which are approximate eigenstates of $S_z$.
$H_z$ tilts the potential and at certain values, levels in opposite wells align, allowing resonant tunneling between wells that results in a marked increase in the
magnetic relaxation
rate.
The
fourth and fifth terms in Eq.~\ref{Mn12tHam} as well as $H_T$
break the commutation of ${\cal H}$ and $S_z$, thereby inducing tunneling.  The tunnel splitting between nearly degenerate states is readily calculated by diagonalizing Eq.~\ref{Mn12tHam}.  The solid lines in Fig.~\ref{ratevHT}b, for $\phi$ = 0 (mod 90$^\circ$), show that destructive interference between tunneling paths induces a dramatic suppression (``quenching") of tunneling at discrete values of $H_T$ for each pair of levels~\footnote{Our calculated tunnel splittings differ somewhat from those presented in Ref.~\onlinecite{645}, even when we use the same Hamiltonian parameters as those reported in that paper.}. (Here and below, we label each state by
$m$, its value of $\left<S_z\right>$ in the absence of tunneling.)  The interference effect is largely destroyed when $\phi$ is increased towards 45$^\circ$ (dashed lines) since $H_T$ then favors one tunneling path
over
others.  The tunneling quenching affects which levels are involved in the magnetic relaxation process, as evidenced by our data.

Crystals of Mn$_{12}$-tBuAc were synthesized according to published procedures~\cite{christos}.  A sample was mounted adjacent to a Hall sensor that was in turn mounted on a rotator probe.  A reference sensor on the same chip was used to measure background signals.
Signals from the two sensors were subtracted with an analog circuit.  We performed extensive measurements on two samples (A \& B).  Measurements of sample B were performed using a modified apparatus (Fig.~\ref{ratevHz} right inset) in which the orientation of both the sample's easy and hard axes relative to the field could be adjusted (the latter \emph{ex situ} by $\sim \pm 35^\circ$).   Measurements were performed as follows.  The sample was rotated
to align
its easy axis with the external magnetic field, magnetizing the sample, i.e. populating one of the wells in Fig.~\ref{ratevHT}a inset.  Next, the sample was rotated to an orientation that produced the desired values of $H_z$ and $H_T$.  The subsequent
time dependence
of the magnetization was monitored
and fit to an exponential decay to extract the
relaxation rate, $\Gamma$. Proceeding this way, we obtained values for $\Gamma$
as a function of $H_z$ and $H_T$ and of temperature $T$ for relaxation near a tunneling resonance.

Some data from sample A are shown in Fig.~\ref{ratevHz}, where $\Gamma$ is plotted as a function of $H_z$ for several values of $H_T$ at $T=$ 3.10 K.  For each value of $H_T$, $\Gamma$ exhibits a roughly Lorentzian dependence on $H_z$, peaked at $H_z=0$, where tunneling is maximum.  $\Gamma$ generally increases with increasing $H_T$ as the tunneling rate is enhanced and, in tandem, the effective energy barrier is reduced~\cite{116}.  We note that for some regions of $H_T$ and $H_z$ (green boxes) the data are bunched -- the relaxation rate changes very little with increasing $H_T$.  The effect is much more pronounced in the shoulders of the peak than near its center.  Fig.~\ref{ratevHT} shows $\Gamma$ as a function of $H_T$ for $H_z=0$ (upper four data sets, from peak center in Fig.~\ref{ratevHz}) and $H_z=-400$ Oe (lower four data sets, from peak shoulders).  All sets  show a roughly exponential increase in $\Gamma$ with $H_T$.  In addition, $\Gamma$ 
exhibits steps and plateaus (the latter corresponding to the bunching in Fig.~\ref{ratevHz}).  These are far more apparent in $H_z=-400$ Oe data.  Each step corresponds to a transition from one dominant pair of tunneling levels (e.g. $m=\pm3$) to another (e.g. $m=\pm4$).  Interestingly, theoretical calculations of these transitions
for
$C=C^\prime=0$ 
predict that the transitions should be independent of $\phi$ and
more pronounced on resonance ($H_z=0$) than
away from resonance~\cite{221, 761}, in contrast to our results.  Structure near $H_z=0$, where $\Gamma$ depends strongly on $H_z$,
may be washed out be inhomogeneous dipole fields.  More importantly, the structure observed near $H_z=-400$  Oe (where  $\Gamma$ is less sensitive to dipole fields) is much more pronounced than predicted by the simple model.  Including
the experimentally determined values of $C$ and $C^\prime$
in Eq.~\ref{Mn12tHam}
induces the tunneling suppression shown in Fig.~\ref{ratevHT}b.   These tunnel quenches, in turn, give rise to the steps in $\Gamma$.

To illustrate this,
Fig.~\ref{ratevHT}b contains a horizontal dotted line -- an empirically determined ``tunnel threshold''.  When the tunnel splitting for a particular pair of levels approaches this threshold, tunneling for that pair begins to become the dominant relaxation mechanism~\cite{497, 116, 221, 145}.  For example, at $H_T \sim 4.2$ kOe (marked by the red vertical dotted line), the tunnel splitting for 
$m=\pm3$ reaches the threshold and tunneling between these levels begins to dominate over relaxation through higher
levels.  This additional relaxation mechanism produces the rapid increase in $\Gamma$ near this field.  Similar transitions occur when other pairs of levels reach the threshold, marked by the other dotted vertical lines in the figure.  The tunneling suppression effect plays a crucial role here: it determines how rapidly the threshold is crossed as $H_T$ increases.  The slope of the tunnel splittings curves in Fig.~\ref{ratevHT}b is rather steep in the vicinity of the threshold right after a quench for $\phi=0$.  In contrast, when $\phi=45^\circ$, the tunnel splitting quenches are absent and the tunnel splittings cross the threshold more gradually, making each transition so broad that it overlaps with others and washing out its observability.

\begin{figure}[tb]
\centering
\includegraphics[width=1\linewidth]{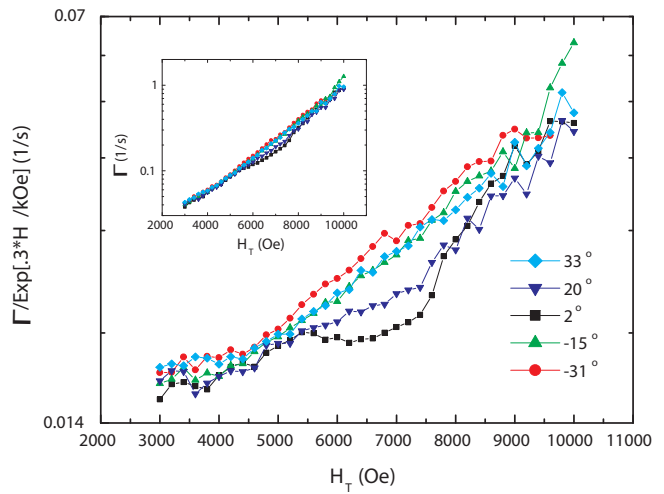}
\caption{(Color online) Relaxation rate from sample B as a function of transverse field for several different values of $\phi^\prime$ for $H_z = -500$ Oe. The inset shows the raw relaxation rate while the main figure shows the same data after dividing by an exponential function to enhance clarity.}
\label{ratevsHT+phi}
\end{figure}

This effect is demonstrated in Fig.~\ref{ratevsHT+phi}. The inset of the figure shows $\Gamma$ as a function of $H_T$ for sample B in the vicinity of one of the transitions for several values of $\phi^\prime = \phi +\phi_{0}$, where $\phi^\prime$ is the experimental azimuthal angle of
the rotator's second stage and $\phi_{0}$ is a constant offset representing the orientation of sample's hard axis (see Fig.~\ref{ratevHz} right inset).
The data features for sample B were less distinct than for sample A, possibly because B was measured more than a year after synthesis.  To enhance clarity, we divided the data by $e^{0.3H_T/\text{kOe}}$, where the coefficient 0.3 was chosen empirically.  The resulting data are presented in the main figure.  For $\phi^\prime=2^\circ$ and $20^\circ$, the sharp transition at $\sim8$ kOe is apparent while it is clearly suppressed for  values of $\phi^\prime$ outside this range.  These results suggest $\phi_{0} \sim9^\circ$, implying that the hard axes for the $m=2$ -- $4$ levels are roughly parallel with the a and b crystallographic axes of the rectangulopiped-shaped crystals~\footnote{In contrast, Ref.~\onlinecite{645} reports that the hard axes are $\sim$45$^\circ$ from the a and b axes.  This apparent discrepancy arises because $C$ and $C^\prime$ have opposite signs, causing the hard-axis directions to depend on which levels $m$ are probed.}.

\begin{figure}[tb]
\centering
\includegraphics[width=1\linewidth]{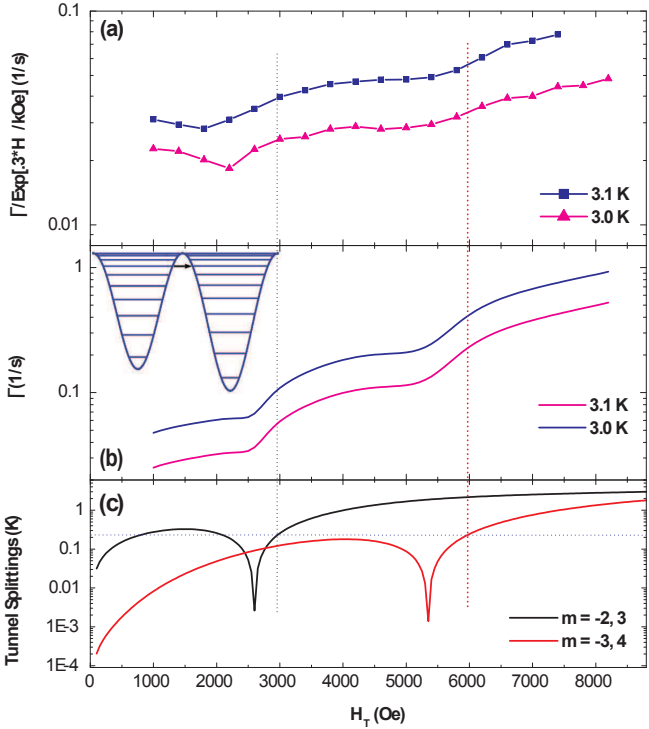}
\caption{(Color online) (a) Relaxation rate (after dividing by an exponential to enhance presentation) as a function of $H_T$ for $H_z$ near the $n=1$ resonance. Data was taken from sample B with experimental azimuthal angle $\phi^\prime=2^\circ$ and $H_z$ set to -1 kOe less than the peak of the resonance. (b) Simulations of relaxation rate as a function of transverse field using the Giant-Spin Hamiltonian model for SMMs with $\phi = 7^\circ$. The inset shows the double-well potential for the $n=1$ resonance. (c) Tunnel splitting calculations with $\phi = 0^\circ$  for the $m$ = -2,3 and $m$ = -3,4 energy-level pairs, as indicated. The dashed horizontal line represents a tunneling threshold while the dashed vertical lines correspond to the transverse fields at which abrupt transitions in relaxation rate occur.}
\label{ratevHTn=1}
\end{figure}

We also studied the relaxation rate near the $n=1$ ($H_z\sim 4.5$ kOe) resonance.
We examined
sample B with $\phi^\prime=2^\circ$ with
$H_z \sim 1$ kOe below the resonance peak.
Again
we see steps in the relaxation rate as a function of $H_T$, as shown in Fig.~\ref{ratevHTn=1}a.  (Like in Fig.~\ref{ratevsHT+phi}, we have removed an exponential background for clarity.) The steps here also correspond to the transitions between dominant tunneling level pairs, as illustrated in Fig.~\ref{ratevHTn=1}c, each occurring when the tunnel splitting for a particular pair rapidly crosses a threshold value (horizontal dotted line) in the wake of a quench.

We performed numerical calculations of $\Gamma$
using a master equation approach~\cite{24,145,447, 221,113, 761} to treat spin-phonon interactions.  For calculational ease, we used the spin's energy eigenbasis, which incorporates tunneling effects automatically since the eigenstates of Eq.~\ref{Mn12tHam} are superpositions of $S_z$ eigenstates.  We neglect off-diagonal elements in the density matrix, a good approximation since our experiments were done away from the exact resonance conditions where such elements are appreciable.  The master equation governing the population of each level, $p_i$, is
\begin{equation}
\frac{{dp_i }}{{dt}} =  \sum\limits_{\scriptstyle j = 1 \hfill \atop
  \scriptstyle i \ne j \hfill}^{21} {-(\gamma _{ij}^{(1)}  + \gamma _{ij}^{(2)}   )} p_i  + {(\gamma _{ji}^{(1) }  + \gamma _{ji}^{(2)}  )} p_j. \label{master}
\end{equation}
The phonon transition rates are given by~\cite{221, 168}
\begin{equation}
\gamma _{ij}^{(\alpha) }  = \frac{{\kappa^{(\alpha)} D^2 }}{{6\pi \rho c_s^5 \hbar ^4 }}\left| s_{ij}^{(\alpha)}
\right|^2
\Delta_{ij}^3 N\left(\Delta_{ij}\right),
\label{rates}
\end{equation}
where $s_{ij}^{(1)}=\left\langle i \right|\left\{ {S_ x  ,S_z } \right\}\left| j \right\rangle$ and $s_{ij}^{(2)}=\left\langle i \right|S_x^2 -S_y^2 \left| j \right\rangle$ and $\Delta_{ij}=\varepsilon_i-\varepsilon_j$, with $\varepsilon_i$ the energy of level $\left| i \right\rangle$.   $N\left(\Delta\right)=(e^{\Delta/k_B T}-1)^{-1}$ is the phonon thermal distribution function, $\rho=1.356\times 10^3$ kg/m$^3$ is the mass density~\cite{645}, and $c_s$ is the transverse speed of sound.  $\kappa^{(1)}=1$ and $\kappa^{(2)}$
is a constant of order unity representing the strength of the associated spin-phonon coupling mechanism~\footnote{
There has been some debate about the magnitude and even the existence of   $\gamma _{ij}^{(2)}$~\cite{205, 207}.  However, we find it necessary to get a reasonable fit to our data.
}.  We neglect possible collective spin-phonon interactions~\cite{763}.
		
We calculated $\Gamma$ by finding the slowest non-zero eigenvalue of the rate matrix implicit in Eq.~\ref{master}.  The calculated rates are fit to the data in Fig.~\ref{ratevHT}, allowing  $c_s$, $\kappa^{(2)}$, $C$ and $C^\prime$ to be unconstrained parameters.  The remaining Hamiltonian parameters were fixed and we set $H_z=-400$ Oe and $\phi=7^\circ$.  The results of the fitting are shown by the solid curves in Fig.~\ref{ratevHT}a.  The calculated rates reproduce the data quite well.  The fit yields $c_s$ = 1122 m/s and $\kappa^{(2)}$ = 1.21.  These parameters set the overall scale of the rate and the general slope of the rate versus $H_T$, respectively.  They do not influence the positions of the steps, which are determined by Hamiltonian parameters $C$ and $C^\prime$.  The fit yields $C=55 ~\mu$K and $C^\prime= -0.81 ~\mu$K, in good agreement with the values determined spectroscopically~\cite{645}.  Using the same parameters, we can also calculate $\Gamma$ for the $n=1$ resonance, shown in Fig.~\ref{ratevHTn=1}b.  Again, the calculations accurately reproduce the structure of the measured relaxation rates.  (Because of the large background for the sample B data (Figs.~\ref{ratevsHT+phi} and \ref{ratevHTn=1}), fits of that data do not produce physically meaningful values of $c_s$ and $\kappa^{(2)}$.)

Our calculations allow us to precisely determine which pairs of levels dominate the tunneling process as a function of $H_T$. The inset of Fig.\ref{ratevHT}b shows an example of the probability ``currents" (dashed arrows)~\cite{145} between some of the relevant states at  $H_T=4.8$ kOe.  The state labels indicate the values of $\left<S_z\right>$.  Diagonal arrows correspond to tunneling transitions in the energy eigenbasis -- such transitions would be forbidden in the absence of tunneling.  For
this
example, $m=\pm3$ are clearly the dominant tunneling levels.
These calculations confirm the interpretation of the steps in the relaxation rate given above, e.g. for $n=0$ at $H_T \sim 4.2$ kOe the dominant tunneling pair switches from $m=\pm2$  to $m=\pm3$.

In conclusion, our measurements
provide the first evidence for a geometric-phase interference effect in a truly four-fold symmetric SMM.  The results also
demonstrate this effect
in the thermally assisted tunneling regime, allowing
identification of which levels dominate the tunneling process. It may be possible to observe similar effects in this system in ground-state tunneling.  Such experiments would require lower temperatures and higher magnetic fields.

\begin{acknowledgments}
We thank M. Bal, D. Garanin and A. Garg  for useful discussions, and acknowledge E. Maradzike, G. Gallo, R. Cann and J. Kubasek for their design and technical contributions to this work. Support for this work was provided by the National Science Foundation under grant nos.~DMR-1006519 and DMR-0449516 and by the Amherst College Dean of Faculty.  Some of the numerical calculations presented were performed on the Amherst College Computing Cluster, funded under NSF grant no.~CNS-0521169.
\end{acknowledgments}

\bibliography{Mn12tBuAc}
\end{document}